\documentclass[prb,twocolumn,superscriptaddress,preprintnumbers,a4paper,amsmath,amssymb,showpacs,floatfix]{revtex4}
\usepackage{graphicx}
\usepackage{graphics}
\usepackage{bm}
\usepackage{dcolumn}

\begin{document}

\preprint{APS/123-QED}

\title{Anomalous spin distribution in the superconducting ferromagnet UCoGe studied by polarized neutron diffraction}

\author{K.~Proke\v{s}}
\email{prokes@helmholtz-berlin.de}\affiliation{Helmholtz-Zentrum
Berlin f\"{u}r Materialien und Energy, Hahn-Meitner Platz 1,
D-14109 Berlin, Germany}

\author{A.~de~Visser}
\affiliation{ Van der Waals-Zeeman Institute, Valckenierstraat 65,
Universiteit van Amsterdam, 1018 XE Amsterdam, The Netherlands
 }%

\author{Y.K.~Huang} \affiliation{ Van der Waals-Zeeman Institute,
Valckenierstraat 65, Universiteit van Amsterdam, 1018 XE
Amsterdam, The Netherlands
 }%

\author{B.~F{\aa}k}
\affiliation{
Commissariat \`{a} l'Energie Atomique, INAC, SPSMS, 38054 Grenoble, France% with \\
}%

\author{E.~Ressouche}
\affiliation{
Commissariat \`{a} l'Energie Atomique, INAC, SPSMS, 38054 Grenoble, France% with \\
}%

\date{\today}
%\preprint{}

\pacs{75.25.-j,74.70.Tx,61.05.fm}
\begin{abstract}
We report a polarized neutron diffraction study conducted to reveal
the nature of the weak ferromagnetic moment in the superconducting
ferromagnet UCoGe. We find that the ordered moment in the normal
phase in low magnetic fields ($B \parallel c$) is predominantly
located at the U atom and has a magnitude of $\sim 0.1~\mu_B$ at
3~T, in agreement with bulk magnetization data. By increasing the
magnetic field the U moment grows to $\sim 0.3 ~\mu _B$ in $12$~T
and most remarkably, induces a substantial moment ($\sim 0.2 ~\mu
_B$) on the Co atom directed antiparallel to the U moment. The
anomalous polarizability of the Co 3$d$ orbitals is unique among
uranium intermetallics and might reflect the proximity to a magnetic
quantum critical point of UCoGe in zero field.

\end{abstract}

%%%%%%%%%%%%%%%%%%%%%%%%%
\maketitle

Recently, UCoGe was identified as a new member of the intriguing
family of superconducting ferromagnets~\cite{Huy-PRL-2007}. In
these metallic ferromagnets superconductivity (SC) is realized
well below the Curie temperature, $T_C$, without expelling
magnetic order, and, even more peculiar, SC and ferromagnetism
(FM) are carried by the same electrons. This is at odds with the
standard BCS theory for phonon-mediated $s$-wave SC, because the
ferromagnetic exchange field is expected to inhibit spin-singlet
Cooper pairing~\cite{Berk-PRL-1966}. The unusual coexistence of SC
and FM therefore calls for an alternative model: critical spin
fluctuations near a magnetic instability provide the mechanism to
pair the electrons in spin-triplet Cooper pairs. The
superconducting ferromagnets discovered until now are
UGe$_2$~\cite{Saxena-Nature-2000}, URhGe~\cite{Aoki-Nature-2001},
UIr~\cite{Akazawa-JPCM-2004} and UCoGe~\cite{Huy-PRL-2007}. FM in
these metals has a strong itinerant character and consequently
these metals can be tuned fairly easily by pressure or magnetic
field to a magnetic quantum critical point and as such are
excellent laboratory systems to investigate spin fluctuation
mediated SC. Magnetically mediated SC is a central theme running
through materials families as diverse as the heavy-fermion
superconductors~\cite{Pfleiderer-RevModPhys-2009}, high-$T_c$ cuprates and the newly-discovered
FeAs-based superconductors~\cite{Kamihara-JAMCHEMSOC-2008}.

\begin{figure}
\includegraphics*[scale=0.30,angle=270]{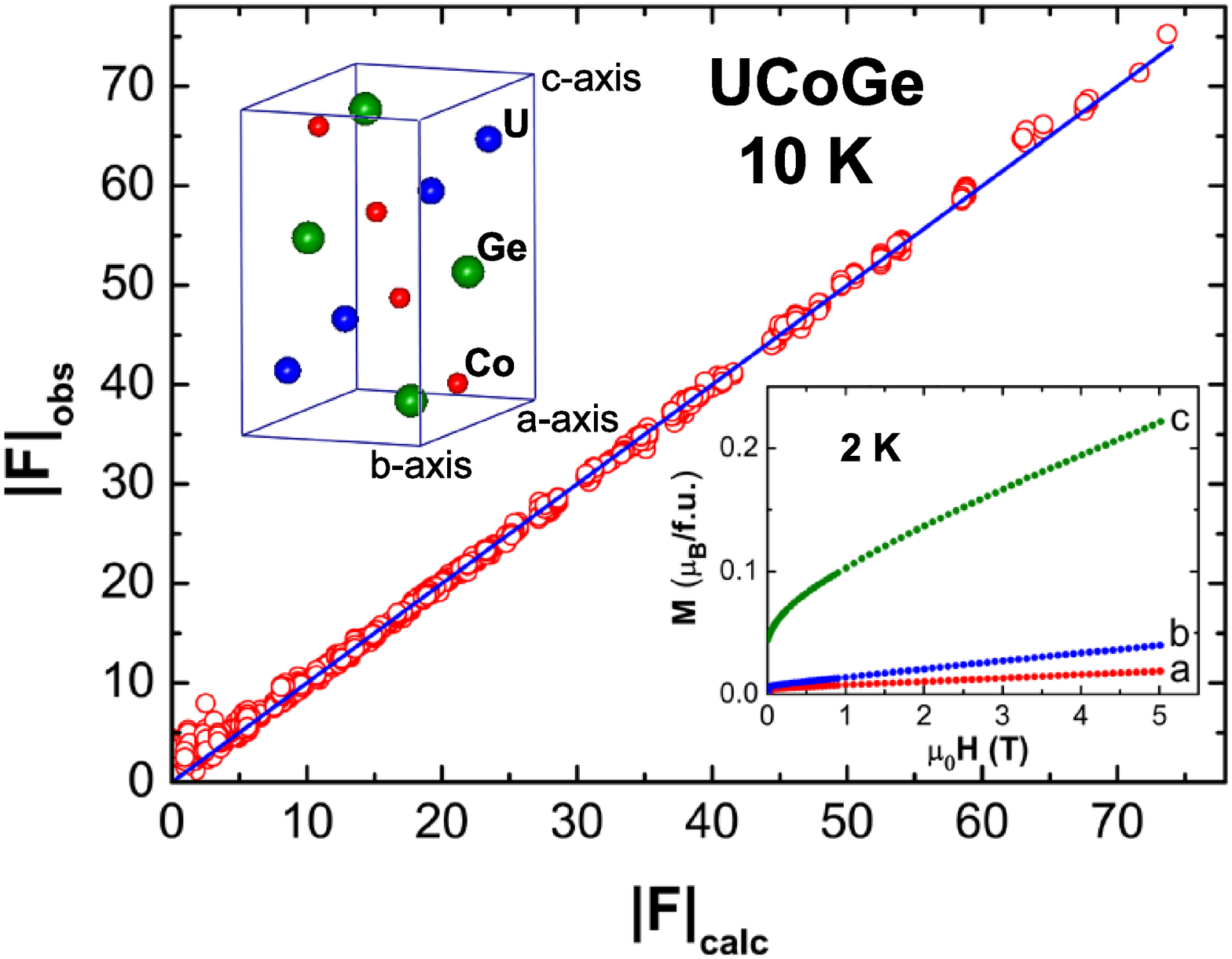}
\caption{(Color online) Observed {\it versus} calculated nuclear
structure factors after correction for absorption and extinctions of
the UCoGe single crystal in the unpolarized neutron diffraction
experiment at $T=10$~K. Upper inset: Schematic representation of the
TiNiSi structure adopted by UCoGe. Lower inset: Magnetization {\it
versus} magnetic field of UCoGe at $T=2$~K for a magnetic field
applied along the three principal axes~\cite{Huy-PRL-2008}.}
\label{Fig1}
\end{figure}

In UCoGe, weak itinerant ferromagnetism develops below the Curie
temperature $T_C = 3$~K~\cite{Huy-PRL-2007}. Magnetization
measurements on single crystals revealed a strong uniaxial
magnetic anisotropy with a small ordered moment $m_0 = 0.07 ~\mu
_B$ directed along the orthorhombic $c$ axis (see inset
Fig.~\ref{Fig1})~\cite{Huy-PRL-2008}. Muon-spin relaxation
experiments provide unambiguous proof for bulk magnetism, which
coexists with SC below the superconducting transition temperature
$T_{sc} = 0.5$~K~\cite{DeVisser-PRL-2009}.

In order to pinpoint the mechanism which gives rise to spin
fluctuation mediated SC in superconducting ferromagnets a detailed
understanding of the magnetic and electronic structure is essential.
In this respect, the polarized neutron diffraction (PND) technique
is an extremely powerful tool as it gives direct information on the
distribution of the magnetization in the unit cell and allows for
the separation of the spin and orbital part of the magnetic
moments~\cite{Nathans-1958-JPhysChemSol}. PND experiments on
UGe$_2$~\cite{Kernavanois-PRB-2001},
URhGe~\cite{Prokes-ActaPhysPol-2003} and
URhSi~\cite{Prokes-PRB-2009} (the latter compounds are isostructural
to UCoGe), show that FM is due to itinerant uranium $5f$ electrons,
and the magnetic moment values are in good agreement with those
derived by electronic structure calculations
~\cite{Shick-PRL-2001,Divis-JAComps-2002}. However, in the case of
UCoGe the discrepancy between magnetization
measurements~\cite{Huy-PRL-2008} and
calculations~\cite{DeLaMora-JPhysCM-2008,Divis-PhysicaB-2008,Samsel-Czekala-JPhysCM-2010}
is large. The calculations predict a small moment $\mu ^U \sim 0.1
~\mu _B$ at the U site due to an almost complete cancellation of the
orbital $\mu_L ^U$ and spin $\mu_S ^U$ magnetic moment. In addition,
a much larger moment $\mu ^ {Co} \sim 0.2$-$0.5 ~\mu _B$ is
predicted at the Co atom. The magnetic moments on the U and Co sites
are expected to orient
parallel~\cite{DeLaMora-JPhysCM-2008,Samsel-Czekala-JPhysCM-2010} or
antiparallel~\cite{Divis-PhysicaB-2008} and consequently it is
argued that the magnetic structure is quite complex and, for
instance, magnetic stripe order~\cite{DeLaMora-JPhysCM-2008} or an
antiferromagnetic spin arrangement~\cite{Divis-PhysicaB-2008} have
been proposed.

In this Letter we report PND experiments on UCoGe conducted to
identify the different contributions to the bulk magnetization in
the normal phase (we apply a field $B
\parallel c$ larger than the upper critical field $B_{c2}^c \simeq 0.5$~T~\cite{Huy-PRL-2008}).
We obtain a surprising result: in low magnetic fields the
magnetization density is predominantly centered at the U atom, but
in a large field (12~T) a substantial Co moment develops, which is
directed antiparallel to the total U moment. The Co moment grows
faster than the U moment. Such a high polarizability of the Co
$3d$ orbitals is highly unusual~\cite{Sechovsky-handbook-1998} and
reflects the proximity to a magnetic quantum critical point of
UCoGe in zero field.

UCoGe crystallizes in the orthorhombic TiNiSi structure with space
group Pnma~\cite{Canepa-JALCOM-1996} (see inset Fig.~1). Neutron
diffraction experiments were carried out on a carefully heat
treated single crystal~\cite{Huy-JMMM-2009}, prepared in a tri-arc
furnace by the Czochralski technique. The sample was shaped into a
bar along the $b$ axis with dimensions $1 \times 1 \times
5$~mm$^3$. Resistivity measurements attest the high quality of the
sample. The residual resistance ratio is 30, $T_C = 2.8$~K and
$T_{sc}=0.6$~K. Magnetization data taken for a field along the
orthorhombic $a$, $b$ and $c$ axis at $T= 2$~K are shown in the
lower inset of Fig.~\ref{Fig1}~\cite{Huy-PRL-2008}. The bulk
magnetic moment at $T=0.1$~K in 3~T and 12~T can be deduced by
extrapolating the magnetization data for $B
\parallel c$ and amounts to 0.17~$\mu_B$/f.u. and
0.35~$\mu_B$/f.u., respectively.

The nuclear structure parameters of the single crystal were
determined at the D15 diffractometer installed at the Institute
Laue-Langevin (ILL) with a wavelength of 1.17~\AA ~in a
four-circle geometry using a closed cycle refrigerator. Absorption
and extinction corrections were made. A large data set comprising
of 1169 reflections was recorded at 10 K. The refinement of the
structure with residual $R_w = 1.2$~\% (see Fig.~\ref{Fig1})
yields lattice parameters $a = 6.813$~\AA, $b = 4.203$~\AA~and $c
= 7.215$~\AA, and atomic coordinates close to those reported in
Ref.~\cite{Canepa-JALCOM-1996}.

In a neutron diffraction experiment on a ferromagnet one typically
measures the magnetic structure factor $F_M({\bf Q}) \propto
\sum_{j} \mu _{j\perp} f_j ({\bf Q}) e^{i\bf{Q} \cdot \bf{r_{j}}}$,
where $\mu _{j\perp }$ is the component of the j-th magnetic moment
perpendicular to the scattering vector $\textbf{Q}$ and
$f_j(\bf{Q})$ is the magnetic form factor of the $j$-th ion at
position $r_{j}$ in the unit cell. Using unpolarized neutrons one
records an intensity proportional to the sum of $|F_M({\bf Q})|^2$
and the nuclear structure factor squared $|F_N({\bf Q})|^2 \propto
|\sum_{j} b_j e^{i\bf{Q} \cdot \bf{r_{j}}}|^2$. However, when the
magnetic moment is small, as is the case for UCoGe, $|F_M({\bf
Q})|^2$ is too small compared to $|F_N({\bf Q})|^2$ and cannot be
determined precisely. A familiar way to improve the sensitivity is
the use of polarized neutrons~\cite{Nathans-1958-JPhysChemSol}. In
the PND experiment one then measures the intensities
 $I^{\pm}(\bf{Q}) \propto$ $|F_N$($\textbf{Q}$) $\pm$
$F _M$($\textbf{Q}$)$|^2$, where the $+$ and $-$ sign refer to up
and down polarization directions of the incoming neutron beam. In
practise one collects flipping ratios $R(\textbf{Q}$)=$I^{+}({\bf
Q})/I^{-}(\bf{Q})$ at many Bragg reflections. The precise
knowledge of $F_N$($\textbf{Q}$) that is determined in the
unpolarized experiment is crucial to evaluate $F_M$($\textbf{Q}$)
and the magnitude of the magnetic moment.

The PND experiment was carried out at the D23 diffractometer at the
ILL with the neutron beam polarized to 92~\%. The UCoGe single
crystal was glued to the cold finger of a dilution refrigerator with
the $c$ axis vertical. Two data sets $R(\textbf{Q}$) were collected
at $T=0.1$~K in magnetic fields of 3~T and 12~T applied along the
easy direction for magnetization ($c$ axis). Each data set consisted
of typically 60 inequivalent reflections of the ($hk0$) and ($hk1$)
type.

\begin{figure}
\includegraphics*[scale=0.33,angle=270]{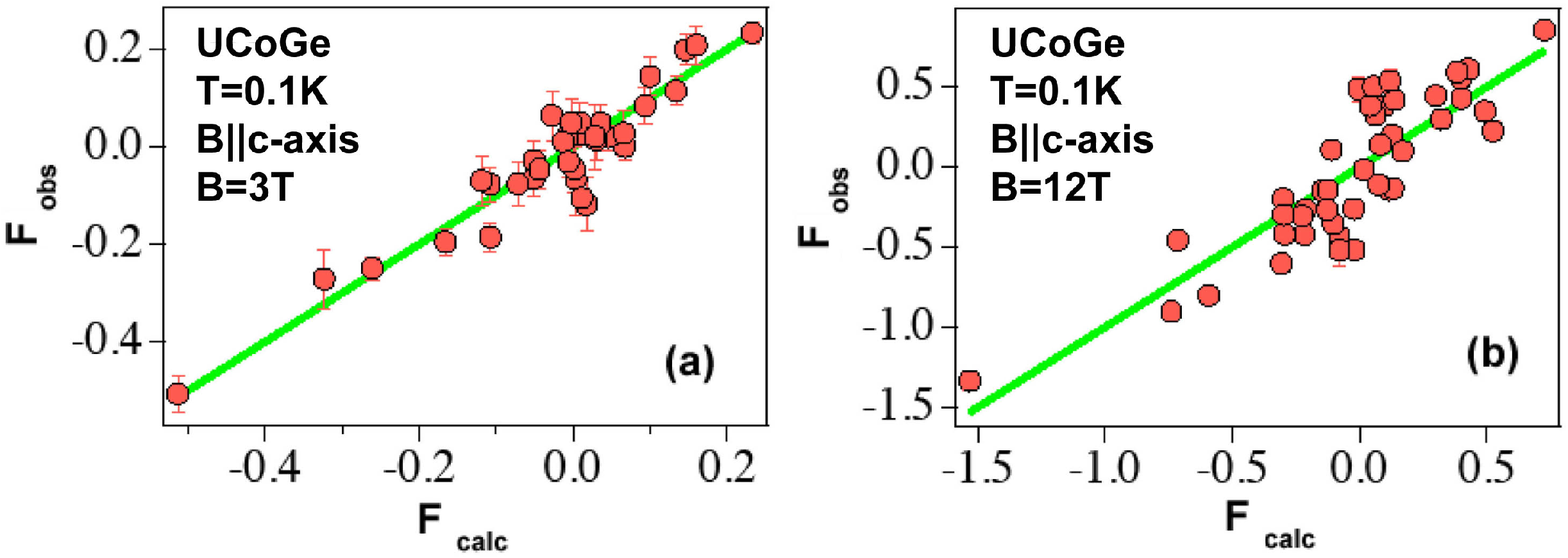}
\caption{(Color online) Observed {\it versus} calculated (solid
line) magnetic structure factor of UCoGe for the polarized neutron
diffraction experiment at $T=0.1$~K in an applied field ($B
\parallel c$) of 3 T (a) and 12 T (b).}
\label{Fig2}
\end{figure}

The uranium magnetic form factor is usually expressed within the
dipolar approximation by the formula $f(\textbf{Q}$) = $\langle
j_0$ ($\textbf{Q}$) $\rangle + C_2 \langle j_2 (\textbf{Q})
\rangle $, where $C_2$ = $\mu _L ^U / ( \mu _S ^U + \mu _L ^U ) =
\mu _L ^U$ / $\mu ^U$ and $j_i$ is the radial integral for the
relevant U$^{3+}$ or U$^{4+}$
configuration~\cite{Brown-book-1992}. An equivalent expression can
be written down for the Co magnetic form factor. By assuming a
magnetic moment on the U or Co site only, we could not obtain a
good fit of the experimental data $F_M$($\textbf{Q}$). However,
when we assume that the U and Co atoms both carry a magnetic
moment the refinement of the magnetic structure (see
Fig.~\ref{Fig2}) leads to a much better fit ($\chi^2$ reduces by a
factor of two and three for the 3 and 12 T data, respectively). In
modeling the form factor we took into account spin and orbital
contributions on the U site, but a spin-only contribution on the
Co site. The fit results did not allow to resolve the uranium
valency because the magnetic form factors of U$^{3+}$ or U$^{4+}$
are very similar. On the other hand, the parameter $C_2$ depends
strongly on the ion state of uranium. The best fits yield the
moment values listed in Table~\ref{tab:table1}. The spin and
orbital moments on the U atoms are antiparallel to each other.
Remarkably, we find a significant spin moment on the Co site,
which is oriented parallel to $\mu _S ^U$, but antiparallel to the
total $\mu ^U$. The obtained values of $C_2$ are close to value
calculated in the intermediate coupling scheme for the free
U$^{4+}$ ions. Obviously, the values of $\mu ^U$ are smaller than
the free ion values. This is in line with the itinerant nature of
the 5$f$ states in UCoGe.

\begin{figure}
\includegraphics*[scale=0.30,angle=0]{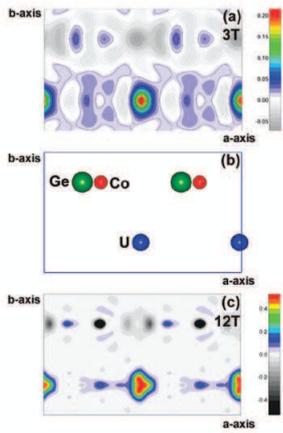}
\caption{(Color online) (a) and (c) Magnetization distribution of
UCoGe obtained form maximum entropy method projected onto the $a$-$b$ plane measured in a field $B
\parallel c$ of 3~T and 12~T, respectively, at $T=0.1$~K; (b)
crystallographic unit cell projected onto the $a$-$b$ plane. In
all cases only half of the unit cell is projected. Notice the
scales (in units $\mu_B$ \AA$^{-2}$) differ in panels (a) and (c).
In the lower panel the density at the Co position is off-scale and
reaches -2.5 $\mu_B$ \AA$^{-2}$.} \label{Fig3}
\end{figure}

Another elegant, powerful and independent treatment of the data is
the method of maximum entropy~\cite{Skilling-book-1985}. This
technique gives the most probable magnetization distribution map
compatible with the measured structure factors and their
experimental uncertainties. Compared to the usual Fourier
synthesis it does not need any {\it a priori} assumptions
concerning the unmeasured Fourier components, which reduces both
the noise and truncation effects. At the same time no detailed
atomic model is needed for the refinement. The basic input
required is the space group, the lattice constants and the
flipping ratio's together with the corresponding measured nuclear
structure factors. The unit cell of UCoGe was divided into $64
\times 64 \times 64 = 262144$ cells, in which the magnetization is
assumed to be constant. The reconstruction was started from a flat
magnetization distribution with a total moment in the unit cell
equal to the bulk magnetization measured experimentally. Our most
important results are summarized in Fig.~\ref{Fig3}, where we have
plotted the resulting magnetization density obtained from the data
collected at 3~T and 12~T ($B
\parallel c$) projected on the $a$-$b$ plane in panel (a) and (c),
respectively. The projected crystal structure is plotted in panel
(b). The density map obtained from the 3~T data set exhibits a
clear, positive density around the uranium position, whereas the
density around the Co position is very small. The 12~T map, however,
is extraordinary: the density at the uranium site has more than
doubled with respect to the 3~T value, but at the same time a
strongly localized, negative density has appeared at the Co site. By
integrating over three dimensions around the U and Co atomic
positions we obtain moments $\mu ^U$ and $\mu ^{Co}$ as listed in Table 1. The values of
$\mu ^U$ are in good agreement with the ones extracted from fitting
the form factor, while the values of $\mu ^{Co}$ are about a factor
3 smaller.

We now have a detailed understanding of the magnetization density on
a microscopic level and proceed to make several important
conclusions. First, we conclude that the weak ferromagnetic state in
UCoGe at low fields is predominantly carried by the U $5f$ moments.
This is at variance with the electronic structure
calculations~\cite{DeLaMora-JPhysCM-2008,Divis-PhysicaB-2008,Samsel-Czekala-JPhysCM-2010}.
However, the PND data reveal that the Co moment is susceptible to a
magnetic field, and magnetic moments on both the U and Co atoms, as
predicted by the calculations, do occur in applied magnetic field.
The small value of the Co moment in weak magnetic fields is in line
with recent zero-field muon spin relaxation
($\mu$SR)~\cite{DeVisser-PRL-2009} and $^{59}$Co Nuclear Quadrupole
Resonance (NQR)~\cite{Ohta-JPhysSocJpn-2010} measurements. Secondly,
in a magnetic field a moment $\mu ^{Co}$ is induced on the Co site,
oriented antiparallel to $\mu ^U$ but parallel to $\mu_S ^U$. While the antiparallel orientation of the spin and orbital $\mu ^U$ parts is common in 5$f$ systems~\cite{Moore-RevModPhys-2009}, the $\mu ^{Co}$ moment is surprisingly large: at 12~T $| \mu ^{Co} / \mu ^U |
\approx 0.4$-$0.8$, depending on the method of analysis. Thus in a
large field $B \parallel c$ the spin arrangement in UCoGe is
ferrimagnetic rather than ferromagnetic. Thirdly, we conclude that
both $\mu ^U$ and $\mu ^{Co}$ grow steadily with increasing
$B\parallel c$. As expected, the magnetic field stabilizes
ferromagnetic order and UCoGe is tuned away from the ferromagnetic
instability. As a fourth important result, we find that the total
magnetic moment $\mu ^U + \mu ^{Co}$ detected in the PND experiment
is lower than the value deduced from the bulk magnetization. This
indicates that the polarization of the interstitial regions and the
contribution from the conduction electrons, which are neglected in
the analysis of the PND data, play an important role in the
magnetization process of UCoGe.

The results of our PND study allow us to draw a close parallel
between UCoGe and URhGe: in low magnetic fields itinerant FM is
predominantly due to the U $5f$ electrons, but the magnetic
interaction strength is different. This offers a unique
opportunity to investigate spin fluctuation mediated SC in a
systematic way. A first step in this direction was recently made
by the extraordinary discovery of field-reentrant SC in
UCoGe~\cite{Aoki-JPhysSocJpn-2009} and
URhGe~\cite{Levy-Science-2005}. Evidence has been presented that
these exotic superconducting states are closely connected to the
enhancement of spin fluctuations associated with a
spin-reorientation process which occurs in high magnetic fields $B
\parallel b$~\cite{Aoki-JPhysSocJpn-2009,Levy-NaturePhysics-2007}.
As concerns UCoGe, for $B \parallel b$ the magnetization is linear
in field and much smaller than for $B \parallel c$ (see
Fig.~\ref{Fig1}) and we do not expect that a moment is induced on
the Co site for this orientation.

Finally, we wish to stress the special role of the $5f$-$3d$
hybridization in UCoGe. In other magnetically ordered orthorhombic
U$TX$ compounds (where $T$ is a transition metal and $X$ is Si or
Ge) no sizeable moments are found on the transition metal
atoms~\cite{Sechovsky-handbook-1998}. This indicates the strong
polarizability of the Co $3d$ orbitals is directly related to the
unique feature of UCoGe, namely the proximity to a magnetic
instability in zero field ~\cite{Huy-PRL-2007}. The application of a
magnetic field drives the system away from the quantum critical
point, which at the same time tends to stabilize $\mu ^U$ and $\mu
^{Co}$. Induced magnetic moments on the transition metal $T$ atom
have also been observed for magnetically ordered hexagonal U$TX$
compounds, like UCoAl~\cite{Javorsky-PRB-2001}. Here the induced
$\mu ^{Co}$ is smaller and the ratio $|\mu ^{Co} / \mu ^U | \approx
0.2$ does not vary with the magnetic field.

In summary, we have conducted polarized neutron diffraction
experiments on a single crystal of the superconducting ferromagnet
UCoGe for $B \geq B_{c2} \parallel c$ in order to solve the nature
of the weak ferromagnetic state. The diffraction data are analyzed
by two different methods: (i) fitting the data to a magnetic form
factor expression with moments on both the U and Co sites, and
(ii) by integrating the magnetization density maps produced by the
maximum entropy method. Both methods reveal that the weak
ferromagnetic magnetic state in small applied magnetic fields is
predominantly due to the U $5f$ moments. However, in a strong
magnetic field a substantial moment on the Co atom is induced,
antiparallel to the U moment, giving rise to a ferrimagnetic spin
arrangement. The unusual polarizability of the Co $3d$ states
points to a strong $5f$-$3d$ hybridization and might provide the
key ingredient to understand the large anisotropy of the upper
critical field $B_{c2}$~\cite{Huy-PRL-2008}.

We acknowledge the ILL for the allocated beamtime and P. Fouilloux
for technical support during the experiment. K.P. acknowledges ILL
for funding of his stay.

\begin{table*}[h]
\caption{\label{tab:table1}Magnetic moment values of UCoGe
determined from the magnetization, $\mu _{bulk}$, compared to the
moments extracted from the PND experiment by the analysis of the
form factor, where the flipping ratios were fitted to a model
allowing for both uranium $\mu ^U$ and cobalt $\mu ^{Co}$ magnetic
moments, and by the integration of the spin density maps obtained by
a maximum-entropy method. The PND experiment was carried out at $T =
0.1$~K in a magnetic field $B \parallel c$ of 3 T and 12 T. We
assumed the uranium moment to have both spin $\mu _S ^U$ and orbital
$\mu _L ^U$ part, whereas the cobalt moment was assumed to have only
a spin part. The parameter $C_2$ = $\mu _L ^U$ / $\mu ^U$ and $\mu
_{int} = \mu _{bulk} - \mu ^U - \mu ^{Co}$, which is the magnetic
moment not associated with a particular atomic position, are listed
as well. All units are $\mu _B$.}

\begin{ruledtabular}
\begin{tabular}{c|c|cccccc|ccc}
  \multicolumn{1}{c}{Field} & \multicolumn{1}{c}{Magnetization} & \multicolumn{6}{c}{Form factor} & \multicolumn{3}{c}{Maximum entropy}
   \\  \hline ~\vspace{-0.1in} \\
   & ~~$\mu _{bulk}$ & $\mu _S ^U$ & $\mu _L ^U$ & $\mu ^U$ & $\mu ^{Co}$& $C_2$ & $\mu _{int}$ ~~~~& $\mu ^U$ & $\mu ^{Co}$ & $\mu _{int}  $
    \\
  3~T~~ & ~0.17(1)~~~ & -0.05(2) & 0.18(1) & 0.13(1) & -0.043(7) & 1.4(2) & 0.08(2) ~~~~& 0.10(1) & 0.00(1) &  0.07(3) \\
  12~T~~ & ~0.35(1)~~~ & -0.17(9) & 0.49(9) & 0.32(7) & -0.27(3) & 1.54(9) & 0.30(10) ~~~~& 0.26(1) & -0.10(1) &  0.19(3)  \\
\end{tabular}
\end{ruledtabular}
\end{table*}

%\bibliography{references}

\end{document}